\documentclass[onecolumn, draftcls, journal]{IEEEtran}
\IEEEoverridecommandlockouts
\usepackage{cite}
\usepackage{cuted}  
\usepackage{amsmath,amssymb,amsfonts}
\usepackage{titlesec}
\usepackage{algorithmic}
\usepackage{graphicx}
\usepackage{textcomp}
\usepackage[font=footnotesize,skip=3pt]{caption}
\usepackage{xcolor}
\usepackage{pgfplots}
\pgfplotsset{compat=1.18}
 
\usepackage{mathrsfs}
\usepackage{amsthm}
\usepackage{enumitem}

\def\BibTeX{{\rm B\kern-.05em{\sc i\kern-.025em b}\kern-.08em
    T\kern-.1667em\lower.7ex\hbox{E}\kern-.125emX}}

\begin{document}

\title{Statistical Channel Model for FSO Systems Assisted by a UAV-Mounted IRS
}
\author{\IEEEauthorblockN{Ferdaous Tarhouni\IEEEauthorrefmark{1}, Vasilis K. Papanikolaou\IEEEauthorrefmark{1}, Laura Cottatellucci\IEEEauthorrefmark{1}, Mohammad-Ali Khalighi\IEEEauthorrefmark{2},
and Robert Schober\IEEEauthorrefmark{1}
 }

 \IEEEauthorblockA{
 \IEEEauthorrefmark{1}Friedrich-Alexander-University Erlangen-Nuremberg (FAU), Erlangen, Germany \\
 \IEEEauthorrefmark{2}Aix-Marseille University, CNRS, Centrale Med, Fresnel Institute, Marseille, France
}
}
\maketitle
\begin{abstract} Integrating optical intelligent reflecting surfaces (IRSs) into aerial platforms, such as unmanned aerial vehicles (UAVs), has been proposed to relax the line-of-sight (LoS) constraint, extend coverage, and enhance deployment flexibility of free space optical (FSO) systems. However, misalignment errors induced by the UAV hovering, in both position and orientation, may degrade connectivity and impair the end-to-end channel quality. In this paper, we develop novel expressions for the electric fields incident on and reflected by an optical IRS, based on the Huygens-Fresnel principle. The resulting expressions are applicable for any combination of incident and reflected propagation directions. Building on this framework, we derive a closed-form statistical channel model that captures the geometric and misalignment losses of an FSO link assisted by a UAV-mounted IRS in the presence of random UAV fluctuations. In particular, we develop a statistical model for the beam misalignment at the receiver lens, assuming Gaussian fluctuations in both the UAV position and orientation. The proposed analytical model is validated through Monte Carlo (MC) simulations and is further used to provide practical design guidelines regarding the optimal placement of the UAV-mounted IRS for the minimization of the outage probability.
\end{abstract}

\section{Introduction}\label{intr}
Free space optical (FSO) communication has been widely adopted in recent years as an effective solution for ``last-mile'' connectivity \cite{8869705} and is considered a strong candidate for 5G and beyond backhaul and fronthaul networks \cite{10821003} owing to its significantly higher data rates compared to radio frequency (RF) communication. 
However, maintaining a line-of-sight (LoS) link between the transmitter and the receiver is a strict requirement for reliable FSO communication. To relax this constraint, alternative solutions such as optical relays \cite{safari2008relay} and optical intelligent reflecting surfaces (IRSs) \cite{najafi2021intelligent}, \cite{ndjiongue2022design} have been proposed. 
Unlike optical relays, which may substantially increase hardware system and signal processing complexity,
optical IRSs consist of planar structures capable of reconfiguring the phase and polarization of incident waves while enhancing the system energy efficiency \cite{ndjiongue2021analysis}. Moreover, IRSs can be mounted on flying platforms such as unmanned aerial vehicles (UAVs) or high-altitude platforms (HAPs) to enhance deployment flexibility, extend coverage, and improve resilience to link failures \cite{11122651}, \cite{11106762}. 

Several works have shed light on IRS-assisted FSO links in aerial-to-ground networks. In \cite{malik2022performance}, the authors study the performance of a UAV-based IRS-assisted RF/FSO network accounting for pointing and phase shift errors for a given IRS phase-shift configuration. In addition, in \cite{sipani2024irs}, the authors examine the impact of UAV hovering at both the IRS and the receiver in terms of angle-of-arrival fluctuations, assuming uniformly distributed angles, and derive statistical channel models based on Gauss-Laguerre approximations for independent incidence and reflection statistics. The authors of \cite{trinh2025optical} explore the use of optical IRSs to support quantum key distribution (QKD) links between HAPs and low-altitude platforms, where the misalignment induced by UAV hovering is studied for different phase shift profiles. In the aforementioned works, the study of the impact of UAV jitter on IRS-assisted FSO systems is based on channel models for the geometric and misalignment loss (GML) derived under restrictive assumptions.  More specifically, these models were developed for a specific phase-shift profile \cite{malik2022performance} or are only valid for in-plane propagation \cite{trinh2025optical}, which limits the deployment flexibility and the applicability of these models to practical scenarios. To the best of the authors’ knowledge, a general statistical channel model for IRS-assisted FSO links that jointly accounts for geometric losses and UAV fluctuations in both position and orientation, and is valid for arbitrary yet fixed transmitter (Tx) and receiver (Rx) locations, is not available in the literature.
Addressing this limitation is essential, as practical deployment scenarios are not limited to in-plane transmission and reflection, especially when aerial platforms are involved. Moreover, tractable statistical channel models are essential for system-level analysis and optimization.

In this paper, we consider a UAV-mounted IRS integrated into an FSO link to connect a Tx with an obstructed Rx. The main contributions are summarized as follows. First, based on the Huygens-Fresnel principle, we derive a novel expression for the electric field incident on the IRS for an arbitrary beam propagation direction from the laser source (LS) along with novel expressions for the incident phase, beam width, and wavefront curvature radius. We also obtain a closed-form expression for the electric field received at the lens for arbitrary Rx orientation, whereas the results in \cite{ajam2022modeling} and \cite{sipani2023modeling} are limited to in-plane propagation. In addition, we derive a tractable approximation of the received power avoiding the upper-bound commonly used in the literature \cite{najafi2021intelligent}, \cite{ajam2023optical}. Second, we develop a mathematical framework to model the UAV-induced misalignment arising from \textit{random} position and angular fluctuations.  In particular, we derive analytical expressions \textit{linking} the incidence and reflection statistics for arbitrary fluctuations. Third, we derive a novel closed-form expression for the probability density function (PDF) of the GML under the considered fluctuation model and for both linear phase shift (LP) and quadratic phase shift (QP) profiles.
To the best of the authors' knowledge, prior works have provided only statistical models assuming a specific phase shift profile \cite{najafi2021intelligent}, \textit{deterministic} received-power expressions \cite{ajam2022modeling}, or Gauss-Laguerre-based approximations for the independent fluctuations \cite{sipani2023modeling}. Based on the proposed framework, we analyze the end-to-end outage probability including GML and atmospheric loss. Finally, simulations validate the accuracy of the analytical results and provide insights into IRS deployment for outage probability minimization.

\section{System and Channel Models}
\subsection{System Model}\label{AA}
\begin{figure}[h]
\centering
\includegraphics[width=0.9
\linewidth]{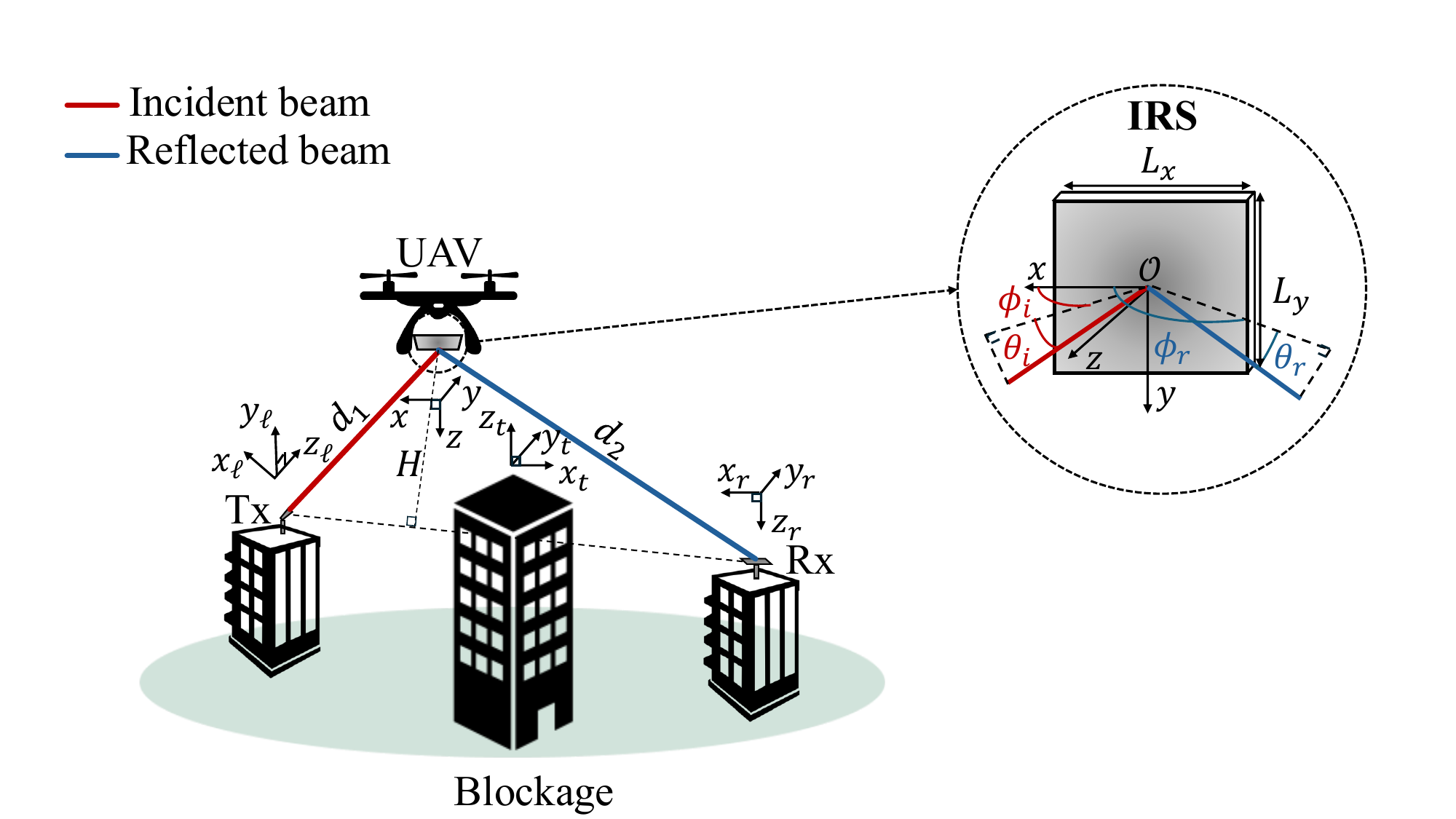}
\caption{FSO system assisted by a UAV-mounted IRS.}
\label{system}
\end{figure}

We consider a UAV-assisted FSO system, as depicted in Fig. \ref{system}, where Tx and Rx communicate via a single UAV-mounted IRS in the absence of a LoS link between the two end points. Tx and Rx are rooftop-mounted and mechanically stable, and equipped with beam tracking and pointing systems, resulting in negligible Tx- and Rx-induced misalignment. In the absence of UAV fluctuations, the IRS is placed at a fixed altitude $H$ above the $x_ty_t$-plane of the $x_ty_tz_t$-coordinate system, which has its origin at the midpoint between the Tx and Rx. This coordinate system will serve later as the reference frame for the numerical search of the optimal IRS location. The IRS is centered at the origin  of the $xyz$-coordinate system, defined in conditions of perfect alignment, lies on the $xy$-plane which is parallel to the $x_ty_t$-plane and the $z$-axis points in the opposite direction of the $z_t$-axis. The IRS sides are parallel to the $x$- and $y$-axes and their dimensions are denoted by $L_x$ and $L_y$, respectively. For analytical tractability, we assume that the UAV and IRS are rigidly connected and they move together along the $z$-axis. We assume a \textit{passive} metamaterial-based optical IRS, modeled as a continuous surface with a continuous phase shift profile, which is justified as $L_x, L_y \!\gg \!\lambda$, where $\lambda$ is the optical wavelength. The Tx comprises an LS at the origin of the $x_\ell y_\ell z_\ell$-coordinate system, emitting a Gaussian laser beam towards the IRS. The $z_\ell$-axis is along the beam axis, the $y_\ell$-axis is parallel to the intersection line of the LS plane and the $xy$-plane. In the absence of beam misalignment, the incident beam intersects the IRS at its center, at distance $d_1$ from the LS, and points in direction $(\theta_i,\phi_i)$, where $\theta_i$ is the angle between the incident beam axis and its projection on the $xy$-plane, and $\phi_i$ is the angle between the $x$-axis and the beam axis projection on $xy$. The Rx is equipped with a photo-detector (PD) and a small circular lens of radius $a$, centered at the origin of the $x_ry_rz_r$-coordinate system at distance $d_2$ from the origin of the $xyz$-coordinate system. The reflected beam points in direction $(\theta_r,\phi_r)$, defined analogously to $(\theta_i,\phi_i)$, and intersects the Rx lens at its center in the absence of beam misalignment. Unlike previous studies \cite{ajam2022modeling}, \cite{10001121}, \cite{najafi2021intelligent}, that assume in-plane beam propagation, we treat the general case of arbitrary Tx and Rx orientations, i.e. $\phi_i\neq0$ and $\phi_r\neq\pi$, which more accurately captures realistic deployment scenarios. The $x_ry_rz_r$-coordinate system is obtained by translating the $xyz$-coordinate system, in the absence of UAV fluctuations, such that the origin of the new coordinate system has coordinates $\mathbf{t}\!=\!(t_x,t_y,t_z)$, where $t_x$, $t_y$, and $t_z$ denote the coordinates along the $x$-, $y$- and $z$-directions, respectively. This translation enables the lens to receive beams from arbitrary directions.

We assume that the UAV fluctuates in both position and orientation due to dynamic wind load and internal vibrations \cite{najafi2020statistical}. Since the IRS is large enough to neglect fluctuations along $x$ and $y$, only fluctuations along $z$ are considered, similarly to \cite{najafi2021intelligent}. Angles $\theta_i$ and $\phi_i$ are modeled as random variables (RVs) to capture angular perturbations. The reflection angles, denoted by $\theta_r$ and $\phi_r$, are also RVs whose statistics are derived from those of the incidence angles. We denote by $\mathbf{p}\!=\!(p_x,p_y,p_z)$ the UAV position, with respect to (w.r.t.) the $x_ry_rz_r$-coordinate system, where only $p_z$ is modeled as an RV.

\subsection{Channel Model}
We consider an intensity modulation/direct detection (IM/DD) FSO system assisted by a UAV-mounted IRS. The received signal at the PD is given by
$y_s=\sqrt{P_t}hx_s+\nu$, 
where $x_s\!\!\in \mathbb{R^+}$ is the symbol transmitted by the LS with $\mathbb{E}\{x_s^2\}\!=\!1$. Here, $\mathbb{E}\{\cdot\}$ denotes the expectation, $P_t$ is the total transmit power, $h \in \mathbb{R^+}$ denotes the end-to-end channel gain along the Tx-IRS-Rx propagation path, and $\nu \in \mathbb{R}$ is the zero-mean real-valued additive white Gaussian noise (AWGN) with variance $\sigma_{\nu}^2$. 

In this work, we assume that the FSO channel is impaired by atmospheric loss and GML. We neglect the impact of atmospheric turbulence attenuation in this work, as the primary focus is to study the effect of random UAV-induced fluctuations on the overall system performance. Atmospheric turbulence modeling has been widely studied in literature, and the reader may refer to \cite{najafi2021intelligent} for further details. Therefore, the channel gain, $h$, is modeled as follows
$h\!=\!\zeta h_p h_{\mathrm{irs}}$,
where $\zeta$ is the PD responsivity, $h_p$ and $h_{\mathrm{irs}}$ denote the atmospheric loss and the GML, respectively. Specifically, the atmospheric loss $h_p$ is deterministic and characterizes the power loss due to absorption and scattering. It is modeled as $h_p\!=\! 10^{-\frac{\kappa d_{e2e}}{10}}$, where $\kappa$ is the attenuation coefficient and $d_{e2e}\!=\!d_1\!+d_2$ denotes the end-to-end propagation distance.

The GML term $h_{\mathrm{irs}}$ represents the fraction of the transmitted power that is reflected by the UAV-mounted IRS and collected by the PD. It includes both the deterministic geometric loss caused by the divergence of the optical beam over the propagation path and the random misalignment loss of the laser beam due to the angular and vertical position fluctuations of the UAV.
\section{Channel Model for UAV-mounted IRS Assisted Link}\label{AAA}
In this section, we first derive general expressions for the electric field incident on the IRS, $E_{\mathrm{in}}(\mathbf{r})$, and the electric field reflected from the IRS, $E_\mathrm{r}(\mathbf{r}_r)$. Then, we derive the conditional point-to-point channel coefficient, $h_{\mathrm{irs}}$.
\subsection{Electric Field Incident on the IRS}\label{ITH}
Assuming that the waist of the Gaussian laser beam, $w_0$, is larger than the wavelength, $\lambda$, the electric field of the Gaussian beam emitted by the LS and propagating along the $z_\ell$-axis, can be modelled as \cite{ajam2022modeling}
\begin{equation}
E_{\ell}(\mathbf{r}_{\ell}) = \frac{E_0 w_0}{w(z_\ell)} e^{-\frac{x_\ell^2 + y_\ell^2}{w^2(z_{\ell})} - j k \left( z_{\ell} + \frac{x_\ell^2 + y_\ell^2}{2 R(z_{\ell})} \right) - \tan^{-1}\left(\frac{z_{\ell}}{z_R}\right)}, 
\label{eq:Eli}
\end{equation}
where $\mathbf{r}_\ell=(x_\ell, y_\ell, z_\ell)$ is a point in the coordinate system of the LS, $E_0$ is the electric field amplitude at the origin, and $k=\frac{2\pi}{\lambda}$ is the wave number. 
The beamwidth at distance $z_\ell$ and the radius of the wavefront curvature are given by $w(z_\ell)\!=\!w_0\sqrt{1+(z_\ell/z_R)^2}$ and $R(z_\ell)\!=\!z_\ell[1+(z_R/z_\ell)^2]$, respectively, where $z_R\!=\!\pi w_0^2/\lambda$ is the Rayleigh range.

The following lemma provides a general expression for the electric field incident on the IRS, valid for arbitrary incident directions. 

\textit{Lemma 1:} Assuming that $d_1 \gg L_x,\ L_y$, the electric field incident on the UAV-mounted IRS, denoted by $E_\mathrm{in}(\mathbf{r})$, where $\mathbf{r}=(x,y,0)$ represents a point on the IRS plane in the $xyz$-coordinate system, is given by
\begin{align}
E_{\mathrm{in}}(\mathbf{r})
=C_i\xi_{\mathrm{in}}
e^{
- \frac{x^2}{w_x^2(d_1)}
- \frac{y^2}{w_y^2(d_1)}
- \frac{xy}{w_{xy}^2(d_1)}
- j \psi_{\mathrm{in}}(\mathbf{r})
}.
\label{E_in}
\end{align}
Here, the incident phase is given by {\small$\psi_{\mathrm{in}}(\mathbf{r})
=k(d_1\!-\!x \cos(\theta_i)\cos(\phi_i)
+y \cos(\theta_i)\sin(\phi_i)
+\!\frac{x^2}{2 R_x(d_1)}\!
+\!\frac{y^2}{2 R_y(d_1)}\!
+\!\frac{xy}{R_{xy}(d_1)})\!-\!\psi_0$}, with {\small$\small\psi_0\!\!=\!\!\tan^{-1}\left(d_1/z_R\right)$}, {\small$\small\xi_{\mathrm{in}}\!=\!\sqrt{|\sin(\theta_i)|}$}, {\small$\small C_i=\sqrt{\frac{4\eta P_t}{\pi w^2(d_1)}}$}, {\small$\small P_t\!=\!\frac{\pi E_0^2w_0^2}{4\eta}$}, and $\eta$ is the free space impedance. Furthermore, we use {\small$w_x(d_1) = \tfrac{w(d_1)}{\sqrt{\sin^2(\phi_i)+\sin^2(\theta_i)\cos^2(\phi_i)}}$},
{\small$w_y(d_1)\!= \!\!\tfrac{w(d_1)}{\sqrt{\cos^2(\phi_i)+\sin^2(\theta_i)\sin^2(\phi_i)}}$},
{\small$w_{xy}(d_1)\!\!=\!\!\tfrac{w(d_1)}{\sqrt{2\sin(\phi_i)\cos(\phi_i)\cos^2(\theta_i)}}\!$}, 
{\small$\!R_x(d_1)\!\!=\!\!\tfrac{R(d_1)}{\sin^2(\theta_i)\cos^2(\phi_i)+\sin^2(\phi_i)}$}, 
{\small$R_y(d_1)\!\!=\!\!\tfrac{R(d_1)}{\cos^2(\phi_i)+\sin^2(\theta_i)\sin^2(\phi_i)}$}, and
{\small$R_{xy}(d_1) \!= \!\tfrac{R(d_1)}{\sin(\phi_i)\cos(\phi_i)\cos^2(\theta_i)}$}.

\begin{IEEEproof} The proof is given in Appendix A.
\end{IEEEproof}

Eq.~\eqref{E_in} provides a general characterization of the elliptical Gaussian beam on the IRS, characterized by the effective beam widths $w_x$ and $w_y$ along the $x$- and $y$-directions, respectively, and $w_{xy}$ captures the cross-term beam width. To the best of the authors' knowledge, this work is the first to derive general expressions for the incident electric field, the corresponding incident phase, the effective beam widths, and the wavefront curvature radii for arbitrary Tx locations w.r.t. the IRS. In particular, \eqref{E_in} reduces to \cite[Eq.~(7)]{ajam2022modeling} as a special case when $\phi_i=0$.
\subsection{Received Electric Field}\label{E_r}
We consider an arbitrary observation point on the lens plane, represented in the IRS coordinate system by  $\mathbf{r}_o=(x_o,y_o,z_o)$.
The electric field, reflected by the IRS and received at point $\mathbf{r}_o$, is given by \cite{ajam2022modeling}
\begin{align}
E_\mathrm{r}(\mathbf{r}_o)
&= \frac{1}{j\lambda} \iint_{(x,y)\in \mathcal{E}_{\mathrm{irs}}}
\!\frac{\xi_{\mathrm{in}}\xi_q C_i}{\lVert \mathbf{r}_o - \mathbf{r} \rVert}
e^{-j k \lVert \mathbf{r}_o - \mathbf{r} \rVert }
e^{-j \phi_{\mathrm{irs}}(\mathbf{r})}  e^{-\frac{x^2}{w_x^2(d_1)} - \frac{y^2}{w_y^2(d_1)} - \frac{x y}{w_{xy}^2(d_1)}-j \psi_{\mathrm{in}}(\mathbf{r})}\, \mathrm{d}x\,\mathrm{d}y,
\label{E_r}
\end{align}
where $\xi_q=\sqrt{|\sin(\theta_r)|}$ is the efficiency factor of the IRS \cite{ajam2022modeling}, $\mathcal{E}_{\mathrm{irs}}$ is the IRS surface, $\phi_{\mathrm{irs}}(\mathbf{r})$ is the phase shift profile of the IRS centered at the origin, and $\lVert \mathbf{r}_o - \mathbf{r} \rVert$ denotes the propagation distance. Using a second-order Taylor series expansion \cite{ajam2022modeling}, $\lVert \mathbf{r}_o - \mathbf{r} \rVert$ can be approximated as follows
{\begin{align}
\small
\lVert \mathbf{r}_o - \mathbf{r} \rVert
& \approx \lVert \mathbf{r}_o \rVert
- \frac{x x_o + y y_o}{\lVert \mathbf{r}_o \rVert}
+ \frac{x^2 + y^2}{2 \lVert \mathbf{r}_o \rVert}
- \frac{x^2 x_o^2 + y^2 y_o^2}{2 \lVert \mathbf{r}_o \rVert^3}  -\frac{x y x_o y_o}{\lVert \mathbf{r}_o \rVert^3}.
\label{ro}
\end{align}}

We consider a QP profile for the IRS, given by $\phi_{\mathrm{irs}}(x,y)= k \left( \phi_{x} x + \phi_{y} y + \phi_{x^2} x^{2} + \phi_{y^2} y^{2} + \phi_{xy} x y + \phi_{0} \right),$
where the phase-gradient coefficients are designed such that the total phase given by $\Phi_t\!=\!-\psi_{\mathrm{in}}(\mathbf{r})\!-\!\phi_{\mathrm{irs}}(\mathbf{r})\!-\!k\lVert\mathbf{r}_o-\mathbf{r}\rVert$ is zero. By following a similar derivation as in \cite{ajam2022modeling} but for the general case $\phi_i\neq0$ and $\phi_r\neq\pi$, the coefficients are functions of the incidence and reflection angles and the system geometry, and are given by {\small$\phi_x = \cos(\theta_i)\cos(\phi_i) + \cos(\theta_r)\cos(\phi_r),\;
\phi_y = \cos(\theta_r)\sin(\phi_r) - \cos(\theta_i)\sin(\phi_i),\;
\phi_0 = - (d_1 + d_2),\;
\phi_{x^2}=\frac{\cos^2(\theta_r)\cos^2(\phi_r)\!-\!1}{2 d_2}-\frac{1}{2 R_x(d_1)}-\frac{1}{4d_2},\;
\phi_{y^2}=\frac{\cos^2(\theta_r)\sin^2(\phi_r)-1}{2 d_2}-\frac{1}{2 R_y(d_1)}-\frac{1}{4d_2},\;
\phi_{xy}=\frac{\cos^2(\theta_r)\cos(\phi_r)\sin(\phi_r)}{d_2}-\frac{1}{R_{xy}(d_1)}$}. The LP profile is obtained as a special case of the QP profile by setting $\phi_{x^2}=\phi_{y^2}=\phi_{xy}=0$.

In the following theorem, assuming the QP profile, we derive a closed-form expression for the reflected electric field in \eqref{E_r} for an arbitrary point on the lens plane, represented in the $x_ry_rz_r$-coordinate system by $\mathbf{r}_r=(x_r,y_r,0)$.

\textit{Theorem 1:} Assuming that $d_2\!\!\gg\!a$, the electric field emitted by the LS, reflected by the UAV-mounted IRS, and received at the Rx lens, denoted by $E_\mathrm{r}(\mathbf{r}_r)$, is given by 
{\small\begin{align}
E_\mathrm{r}(\mathbf r_r) &=\!\frac{\pi C_\mathrm{r}}{4\sqrt{b_x \tilde b_y}}
\exp\!\left[\!
-\frac{k^2}{4}\bigg(
\frac{\mathcal{X}^2}{b_x}+\frac{\mathcal{Y}^2}{\tilde b_y}\bigg)\!\right]\!
\Bigg[\operatorname{erf}\!\left(
\sqrt{b_x}\frac{L_x}{2}
+\frac{b_{xy}L_y}{8\sqrt{b_x}}
-\frac{jk}{2\sqrt{b_x}}\mathcal{X}\right)\!-\!\operatorname{erf}
\!\left(-\sqrt{b_x}\frac{L_x}{2}
+\frac{b_{xy}L_y}{8\sqrt{b_x}}
-\frac{jk}{2\sqrt{b_x}}\mathcal{X}\right)\!\Bigg]
\notag \\
&\times\Bigg[\!
\operatorname{erf}\!\bigg(
\sqrt{\tilde b_y}\frac{L_y}{2}
-\frac{jk}{2\sqrt{\tilde b_y}}
\mathcal{Y}\bigg)\!-\!\operatorname{erf}\!\bigg(-\!\sqrt{\tilde b_y}\frac{L_y}{2}
-\frac{jk}{2\sqrt{\tilde b_y}}
\mathcal{Y}\bigg)\Bigg].
\label{Er}
\end{align}}
In \eqref{Er}, $\operatorname{erf}(\cdot)$ denotes the error function and the employed variables and coefficients depend on the effective beam parameters and the geometric coefficients determined by the system configuration as follows: {\small$\mathcal{X}\!=\!c_3 x_r-c_4 y_r$}, {\small$\mathcal{Y}=
\!c_5 x_r+c_6 y_r-\frac{b_{xy}(c_3 x_r - c_4 y_r)}{2b_x}$}, {\small$C_\mathrm{r}\!=\!\frac{C_i\xi_{\mathrm{in}}\xi_q}{j\lambda d_2}
e^{j\psi_0}$}, {\small$b_x=\frac{1}{w_x^2(d_1)}+ j k\left(
\frac{1-c_1^2d_2^2}{2d_2}
+\frac{1}{2R_x(d_1)}
+\phi_{x^2}\right)$}, {\small$b_y=\frac{1}{w_y^2(d_1)}
+j k\left(\frac{1-c_2^2d_2^2}{2d_2}
+\frac{1}{2R_y(d_1)}
+\phi_{y^2}\right)$, $b_{xy}=\frac{1}{w_{xy}^2(d_1)}+jk\left(
\frac{1}{R_{xy}(d_1)}
-c_1c_2d_2+\phi_{xy}
\right)$}, {\small$\tilde b_y\!=\!b_y-\!\frac{b_{xy}^2}{4b_x}$}, {\small$c_1=\frac{\cos(\phi_r)\cos(\theta_r)}{d_2}$}, {\small$c_2=\frac{\sin(\phi_r)\cos(\theta_r)}{d_2}$}, {\small$c_3=\frac{\cos(\phi_r)\sin(\theta_r)}{d_2}$}, {\small$c_4=\frac{\sin(\phi_r)}{d_2}$}, {\small$c_5=\frac{\sin(\phi_r)\sin(\theta_r)}{d_2}$}, and {\small$c_6=\frac{\cos(\phi_r)}{d_2}$}.
\begin{IEEEproof}
The proof is given in Appendix B.
\end{IEEEproof}
Equation~\eqref{Er} provides a tractable closed-form expression that is valid for general incident and reflected beam directions. In particular, \eqref{Er} has a similar mathematical structure as \cite[Eq.~(15)]{ajam2022modeling} and \cite[Eq.~(16)]{sipani2023modeling}. However, these expressions correspond to a special case of \eqref{Er}, when $\phi_i=0$.
\subsection{Conditional GML}\label{CGML}
In this subsection, we derive a closed-form expression for the conditional GML coefficient, $h_\mathrm{irs}$, for a given $u=\lVert \mathbf{b}_r\rVert$, where $\mathbf{b}_r=(b_{r,x},b_{r,y},b_{r,z})$ denotes the center of the beam footprint on the Rx lens, in the $x_ry_rz_r$-coordinate system. Specifically, $u$ represents the distance between the center of the lens and the center of the received beam footprint on the lens, i.e., the misalignment. The deterministic GML coefficient of the UAV-mounted IRS-assisted FSO link is given by \cite{ajam2023optical} 
\begin{equation}
h_{\mathrm{irs}}
= \frac{1}{2 \eta P_t}
\iint_{\mathcal{A}_r} \big| E_{\mathrm{r}}(\mathbf{r}_r) \big|^2 \, \mathrm{d}\mathcal{A}_r,
\end{equation}
where $E_{\mathrm{r}}(\mathbf{r}_r)$ is the received electric field given by \eqref{Er} and $\mathcal{A}_r$ is the aperture of the lens of the PD.

\textit{Lemma 2:} For a practical receiver lens, $\frac{2a}{d_2} \ll 1$ holds. 
We obtain $h_\mathrm{irs}$ in terms of misalignment $u$ as follows
{\small\begin{align}
h_{\mathrm{irs}}
&=\!\sqrt{\frac{\pi}{\rho_y}}\frac{|\tilde C|^2 |C_1|^2|C_2|^2}{4\eta P_t}\!
\int_{-\tilde{a}}^{\tilde{a}}\!
\!\bigg[
\operatorname{erf}\left(\!
\sqrt{\rho_y}\,\tilde{a}+ \frac{\rho_{xy}}{2 \sqrt{\rho_y}}(x_r-u)\right)\!-\operatorname{erf}\left(\!
-\sqrt{\rho_y}\,\tilde{a} + \frac{\rho_{xy}}{2 \sqrt{\rho_y}}(x_r-u)\right)\bigg]e^{-\tilde \rho_x(x_r-u)^2}\mathrm{d}x_r,
\label{hirs1}
\end{align}}
where {\small$\tilde C \!=\! \pi C_\mathrm{r}/(4\sqrt{b_x\tilde b_y})$}. Here, {\small$C_1 \!=\!\operatorname{erf}(\sqrt{b_x}L_x/2+\alpha_1)-\operatorname{erf}(-\sqrt{b_x}L_x/2+\alpha_1)$} with {\small$\alpha_1\!=\![b_{xy}L_y-2jka(c_3-c_4)]/(8\sqrt{b_x})$}, {\small$C_2\!=\!\operatorname{erf}(\sqrt{\tilde b_y}L_y/2-\alpha_2)-\operatorname{erf}(-\sqrt{\tilde b_y}L_y/2-\alpha_2)$} with {\small$\alpha_2\!=\![jka/(4\sqrt{\tilde b_y})](c_5+c_6-\tfrac{b_{xy}}{2b_x}(c_3-c_4))$}, {\small$\tilde{a}\!=\!a\sqrt{\pi}/2$}, {\small$\tilde{\rho}_x\!=\!\rho_x-\rho_{xy}^2/(4\rho_y)$}, {\small$\rho_x\!=\!\tfrac{k^2}{4}[\mathcal{B} c_3^2+\tfrac{1}{\tilde b_y}(\tfrac{b_{xy}}{2b_x}c_3-c_5)^{2}+\tfrac{1}{\tilde b_y^*}(\tfrac{b_{xy}^*}{2b_x^*}c_3-c_5)^{2}]$}, {\small$\rho_y\!=\!\tfrac{k^2}{4}[\mathcal{B} c_4^2+\tfrac{1}{\tilde b_y}(c_6+\tfrac{b_{xy}}{2b_x}c_4)^{2}+\tfrac{1}{\tilde b_y^*}(c_6+\tfrac{b_{xy}^*}{2b_x^*}c_4)^{2}]$}, {\small$\rho_{xy}\!=\!-\tfrac{k^2}{2}[c_3c_4\mathcal{B}+\tfrac{1}{\tilde b_y}(\tfrac{b_{xy}}{2b_x}c_3-c_5)(c_6+\tfrac{b_{xy}}{2b_x}c_4)+\tfrac{1}{\tilde b_y^*}(\tfrac{b_{xy}^*}{2b_x^*}c_3-c_5)(c_6+\tfrac{b_{xy}^*}{2b_x^*}c_4)]$} with {\small$\mathcal{B}\!=\!1/b_x+1/b_x^*$}.

\begin{IEEEproof}  We use the approximations $x_r \approx \frac{a}{2}$ and $y_r \approx \frac{a}{2}$ in the $\operatorname{erf}(\cdot)$ terms of $E_{\mathrm{r}}(\mathbf{r}_r)$ and \cite[Eq.~(2.33-1)]{gradshteyn2014table} to solve the inner integral in $h_\mathrm{irs}$ w.r.t. $y_r$. Then, due to the small size of the circular lens, we can approximate it by a square lens with the same area and side length $a\sqrt{\pi}$ \cite{ajam2022modeling}.
\end{IEEEproof}
Notably, \eqref{hirs1} characterizes the GML, which includes both the deterministic geometric loss and the random UAV hovering-induced misalignment, through variable $u$. It highlights the dependence of the channel coefficient on key system parameters, including the IRS and lens dimensions, the phase-shift profile, and the positions of the LS and the PD. This expression is valid for any location of the LS and PD. Furthermore, it can be evaluated with low computational complexity, as it involves only a one-dimensional integral.

In the following theorem, \eqref{hirs1} is further simplified into a closed-form expression, which will be useful for the subsequent analysis and for deriving the closed-form GML distribution in Section~\ref{stm}.

\textit{Theorem 2:} For practical receiver lens sizes and misalignments, $\frac{|\tilde a|-u}{d_2}\!\ll\!1$ holds. Then, $h_\mathrm{irs}$ is given by
{\small
\begin{align}
&h_{\mathrm{irs}}
=\!\frac{\pi|\tilde{C}|^{2}|C_{1}|^{2}|C_{2}|^{2}\!}
{4\eta P_t\sqrt{\!\rho_y\tilde{\rho}_x}}
\operatorname{erf}\left(\sqrt{\tilde{\rho}_x}\tilde{a}\right)\!\Bigg[\!
\operatorname{erf}\!\left(\sqrt{\rho_y}\tilde{a}+\frac{\rho_{xy}a}{4\sqrt{\rho_y}}\right)\! -\!\operatorname{erf}\!\left(\!-\sqrt{\rho_y}\tilde{a}+\frac{\rho_{xy}a}{4\sqrt{\rho_y}}\right)\!
\Bigg]\!\exp\!\left[-\frac{2\tilde{a}^{2}\left(\tilde{\rho}_x\right)^{\frac{3}{2}}}
{\sqrt{\pi}\operatorname{erf}\left(\sqrt{\tilde{\rho}_x}\tilde{a}\right)}
e^{-\tilde{\rho}_x\tilde{a}^{2}}\!u^{2}\!
\right].
\label{hirs2}
\end{align}
}
\begin{IEEEproof} The proof is given in Appendix C.
\end{IEEEproof}
As a special case, for $\phi_i\!=\!0$, $\phi_r\!=\!\pi$, and $u\!=\!0$, \eqref{hirs2} reduces to \cite[Eq.~(20)]{ajam2022modeling}.
\section{Statistical Model for the GML}\label{stm}
In this section, we derive a statistical model for the GML induced by the UAV hovering. To this end, we first model the position and angular fluctuations of the UAV, and subsequently derive a closed-form expression for the PDF of the GML.
\subsection{Misalignment Model}\label{misal}
The orientation of the UAV and its displacement along the vertical axis are subject to random fluctuations over time. To model these perturbations, we define the following independent RVs, $p_z\!=\!\mu_z+\epsilon_z$, $\theta_i\!=\!\mu_{\theta_i}+\epsilon_{\theta_i}$, and $\phi_i\!=\!\mu_{\phi_i}+\epsilon_{\phi_i}$, where $\mu_s$ denotes the mean, and $\epsilon_s\!\sim\!\mathcal{N}(0, \sigma_s^{2})$ is a zero-mean Gaussian RV with variance $\sigma_s^2$ for $s\!\in\!\{z,\theta_i,\phi_i\}$. Given the position of the LS, represented by $(x_{\mathrm{LS}},y_{\mathrm{LS}},z_{\mathrm{LS}})$, in the IRS-coordinate system, the center of the incident beam footprint on the IRS, in the same coordinate system, is given by $\mathbf{b}_i\!=\!(b_{i,x},b_{i,y},b_{i,z})=\!
\left(x_{\mathrm{LS}}+z_{\mathrm{LS}} \cot(\theta_i)\cos(\phi_i), 
y_{\mathrm{LS}}-z_{\mathrm{LS}} \cot(\theta_i)\sin(\phi_i),
0\right)$. The mean angles $\mu_{\theta_i}$ and $\mu_{\phi_i}$ are determined to ensure that the incident beam line intersects the IRS on its center, i.e., $\mathbb{E}\{\mathbf{b}_i\}=0$. This yields
\begin{subequations}\label{angli}
\begin{align}
\mu_{\theta_i} 
&= -\pi/2 
+ \cos^{-1}\!\left(
z_{\mathrm{LS}}/
\sqrt{x_{\mathrm{LS}}^2 + y_{\mathrm{LS}}^2 + z_{\mathrm{LS}}^2}
\right),\label{angli:a}\\
\mu_{\phi_i} 
&= -\tan^{-1}\!\left(y_{\mathrm{LS}}/x_{\mathrm{LS}}\right).
\label{angli:b}
\end{align}
\end{subequations}
Next, given the $\theta_i$ and $\phi_i$ statistics, we derive the statistics of the reflection angles $\theta_r\!=\!\mu_{\theta_r}+\epsilon_{\theta_r}$ and $\phi_r\!=\!\mu_{\phi_r}+\epsilon_{\phi_r}$, where $\mu_{\theta_r}$ and $\mu_{\phi_r}$ denote the mean angles, and $\epsilon_{\theta_r}$ and $\epsilon_{\phi_r}$ represent the fluctuations, respectively. We express $\mathbf{b}_r$, see Section \ref{CGML}, in terms of the center of the incident beam footprint $\mathbf{b}_i$, as follows $\mathbf{b}_r\!=\!(b_{i,x}-t_x-\cot(\theta_r)\cos(\phi_r)\ p_z,b_{i,y}-t_y+\cot(\theta_r)\sin(\phi_r)\ p_z,0)$. Under the assumption of perfect alignment, the reflected beam line from the IRS intersects the Rx lens on its center and the mean angles $\mu_{\theta_r}$ and $\mu_{\phi_r}$ are determined by enforcing the condition $\mathbb{E}\{\mathbf{b}_r\}\!=\!0$, yielding
\begin{subequations}\label{anglr}
\begin{align}
\mu_{\theta_r}
&= -\pi/2
+ \cos^{-1}\!\left(
\mu_z/\sqrt{\tilde b_{i,x}^2+\tilde b_{i,y}^2+\mu_z^2}
\right), \label{anglr:a}\\
\mu_{\phi_r}
&= -\tan^{-1}\!\left(\tilde b_{i,y}/\tilde b_{i,x}\right),
\label{anglr:b}
\end{align}
\end{subequations}
where $\tilde b_{i,x}=x_{\mathrm{LS}}-t_x+z_{\mathrm{LS}} \cot(\mu_{\theta_i})\cos(\mu_{\phi_i}) \, \text{and} \, \tilde b_{i,y}=y_{\mathrm{LS}}-t_y-z_{\mathrm{LS}} \cot(\mu_{\theta_i})\sin(\mu_{\phi_i})$. Subsequently, for small fluctuations $\epsilon_{\theta_i}$ and $\epsilon_{\phi_i}$, first-order Taylor series expansions of trigonometric functions are used to derive tractable zero-mean Gaussian approximations of $\epsilon_{\theta_r}$ and $\epsilon_{\phi_r}$ in terms of the statistics of $\theta_i$ and $\phi_i$, with $\epsilon_{\theta_r}$ and $\epsilon_{\phi_r}$ being correlated. 
 We obtain $\epsilon_{\theta_r}\!=\!\nu_1\epsilon_{\theta_i}\!+\!\nu_2\epsilon_{\phi_i}$ and $\epsilon_{\phi_r}\!=\!\nu_3\epsilon_{\theta_i}\!+\!\nu_4\epsilon_{\phi_i}$, where, using the shorthands {\small$\tilde x\!=\!(x_{\mathrm{LS}}-t_x)/z_{\mathrm{LS}}$}, {\small$\tilde y\!=\!(y_{\mathrm{LS}}-t_y)/z_{\mathrm{LS}}$}, and {\small$\Delta\!=\!e_5^2+e_6^2$}, we have {\small$\nu_1\!=\!e_3 e_4/(e_1\sqrt{1-e_4^2})$}, {\small$\nu_2\!=\!e_2 e_4/(e_1\sqrt{1-e_4^2})$}, {\small$\nu_3\!=\!(e_5\cos(\mu_{\phi_i})+e_6\sin(\mu_{\phi_i}))/\Delta$}, {\small$\nu_4\!=\!\cos(2\mu_{\theta_i})(e_5\sin(\mu_{\phi_i})-e_6\cos(\mu_{\phi_i}))/(2\Delta)$}, {\small$e_1\!=\!\mu_z^2\sin^4(\mu_{\theta_i})/z_{\mathrm{LS}}^2+(\tilde x\sin^2(\mu_{\theta_i})\!+\!\cos(2\mu_{\theta_i})\cos(\mu_{\phi_i})/2)^2+(\tilde y\sin^2(\mu_{\theta_i})\!-\!\cos(2\mu_{\theta_i})\sin(\mu_{\phi_i})/2)^2$}, {\small$e_2\!=\!\sin^2(\mu_{\theta_i})\cos(2\mu_{\theta_i})(\tilde x\sin(\mu_{\phi_i})\!+\!\tilde y\cos(\mu_{\phi_i}))$}, {\small$e_3\!=\!2\sin^2(\mu_{\theta_i})(\tilde x\cos(\mu_{\phi_i})+\tilde y\sin(\mu_{\phi_i}))+\cos(2\mu_{\theta_i})\cos(2\mu_{\phi_i})$}, {\small$e_4\!=\!-\mu_z\sin(\mu_{\theta_i})/(\sqrt{e_1}|z_{\mathrm{LS}}|)$}, {\small$e_5\!=\!-\tilde y\cos(2\mu_{\theta_i})\tan(\mu_{\theta_i})/2+\sin(\mu_{\phi_i})\cos(2\mu_{\theta_i})/2$}, and {\small$e_6\!=\!-\tilde x\cos(2\mu_{\theta_i})\tan(\mu_{\theta_i})/2-\cos(\mu_{\phi_i})\cos(2\mu_{\theta_i})/2$}.
Finally, writing $\mathbf{b}_r$ directly in terms of the UAV-mounted IRS position $\mathbf{p}$ and the reflection angles $\theta_r$ and $\phi_r$, we obtain \begin{equation}
\mathbf{b}_r
\!=\!(p_x\!- \cot(\theta_r)\cos(\phi_r)p_z, p_y \!+\cot(\theta_r)\sin(\phi_r)p_z,0),
\label{br}
\end{equation}
where $p_x\!=\!-d_2 \cos(\mu_{\theta_r})\cos(\mu_{\phi_r})$, $p_y\!=\!d_2 \cos(\mu_{\theta_r})\sin(\mu_{\phi_r})$, and $p_z\!=\!-d_2 \sin(\mu_{\theta_r})\!+\!\epsilon_z$. 
Recalling that $ u\!=\!\lVert \mathbf{b}_r\lVert=\sqrt{b_{r,x}^2+b_{r,y}^2}$ and using \eqref{br}, we can prove that $u$ follows a Beckmann distribution, as shown in Appendix D.
\subsection{PDF of GML}\label{PDFh}
Using \eqref{hirs2}, we show that the PDFs of $h_\mathrm{irs}$ and $u^2$, denoted respectively by $f_{h_\mathrm{irs}}(\cdot)$ and $f_{u^2}(\cdot)$, are related as follows
\begin{align} 
f_{h_{\mathrm{irs}}}(h) &\!=\! \frac{1}{\beta_{\mathrm{g}}h}f_{u^2}\left(\frac{1}{\beta_{\mathrm{g}}}\ln\left(\frac{\alpha_{\mathrm{g}}}{h}\right)\right), \quad \, 0 < h \le \alpha_{\mathrm{g}}, 
\label{relpdfs} \end{align}
where $\alpha_{\mathrm{g}}\!=\!\frac{\pi|\tilde{C}|^{2}|C_{1}|^{2}|C_{2}|^{2}\!}
{4\eta P_t\sqrt{\!\rho_y\tilde{\rho}_x}}
\operatorname{erf}\left(\sqrt{\tilde{\rho}_x}\tilde{a}\right)\!\bigg[\!
\operatorname{erf}\!\left(\sqrt{\rho_y}\tilde{a}+\frac{\rho_{xy}a}{4\sqrt{\rho_y}}\right)\!-\!\operatorname{erf}\!\left(\!-\sqrt{\rho_y}\tilde{a}+\frac{\rho_{xy}a}{4\sqrt{\rho_y}}\right)\!
\bigg]$ and $\beta_{\mathrm{g}}
\!=\!\frac{2\tilde{a}^{2}\left(\tilde{\rho}_x\right)^{\frac{3}{2}}}
{\sqrt{\pi}\operatorname{erf}\left(\sqrt{\tilde{\rho}_x}\tilde{a}\right)}
e^{-\tilde{\rho}_x\tilde{a}^{2}}$.

The following theorem provides a closed-form statistical channel model for the GML.

\textit{Theorem 3:} Assuming small fluctuations, $\!\epsilon_s$, $s\!\in\!\{\theta_i, \phi_i, \theta_r, \phi_r, z\}$, the PDF of $h_\mathrm{irs}$ is given by
{\begin{equation} 
f_{h_{\mathrm{irs}}}(h) \!=\! \frac{1}{\Gamma(n)\alpha_{\mathrm{g}} m^{n}\beta_{\mathrm{g}}^{n}}\!\!\left(\frac{h}{\alpha_{\mathrm{g}}} \right)^{\frac{1}{m\beta_{\mathrm{g}}}-1}\!\! \left[\ln\!\left(\frac{\alpha_{\mathrm{g}}}{h}\right) \right]^{n-1}, \, \quad 0 \!<\! h \!\le \!\alpha_{\mathrm{g}},
\label{pdfh} \end{equation}}
\par \noindent where $m=
{\operatorname{Var}(u^2)}/{\mathbb{E}\{u^2\}}$, $n
={\mathbb{E}\{u^2\}^2}/{\operatorname{Var}(u^2)}$, {$
\operatorname{Var}(u^2)
=\mathbb{E}\{u^4\} - \mathbb{E}\{u^2\}^2$}, and $\Gamma(\cdot)$ denotes the Gamma function.
\begin{IEEEproof} The proof is given in Appendix D.
\end{IEEEproof}
Unlike the Gauss-Laguerre-based channel model in \cite{sipani2023modeling} or the closed-form PDFs derived for non-IRS FSO links with simpler misalignment distributions in \cite{najafi2020statistical}, \eqref{pdfh} is a closed-form PDF for an FSO-based link assisted by a UAV-mounted IRS, holds for arbitrary transmitter and receiver locations and Beckmann-distributed misalignment. This makes it applicable to more realistic deployment scenarios.

\section{Simulation Results}\label{simul}
In this section, we first verify the accuracy of the derived statistical GML channel model. Then, we provide insights on the UAV-mounted IRS deployment for improving the end-to-end system performance. Unless stated otherwise, the simulation parameters are set as follows: $w_0=3 \ \text{mm}$, $\lambda=1550 \ \text{nm}$, $P_t\!=\!0.4 \ \text{W}$, $a\!=\!15 \ \text{cm}$, $\eta\!=\! 377 \ \Omega$, $H\!=\!100 \ \text{m}$, $\zeta\!=\!0.5$, and $\kappa\!=\!0.43 \times 10^{-3} \ \frac{\text{dB}}{\text{m}}$ \cite{ajam2022modeling}, \cite{ajam2023optical}. We consider a square IRS with side length $L_x\!=\!L_y=0.5 \ \text{m}$, the LS is placed at $(x_{\mathrm{LS}},y_{\mathrm{LS}},z_{\mathrm{LS}})=(500, -100, 100) \ \text{m}$ w.r.t. the $xyz$-coordinate system, and $\mathbf{t}$ is given by $(-500 , 100, 100) \ \text{m}$. All simulation results are based on $10^7$ Monte Carlo (MC) realizations of RVs $\epsilon_s$ for $s\!\in\!\{z,\theta_i,\phi_i\}$.

\begin{figure}[h]
\centering
\includegraphics[width=0.75
\linewidth]{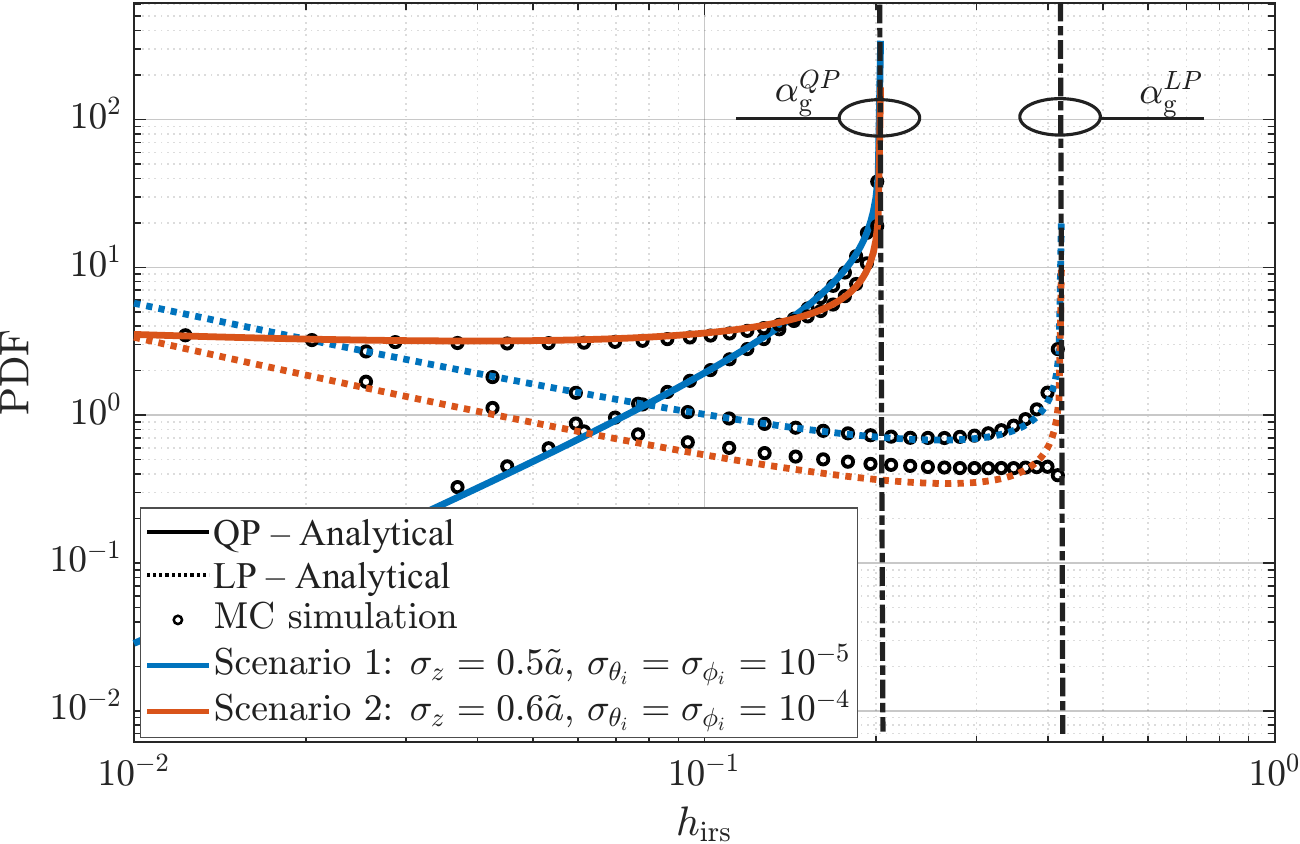}
\caption{PDF of the GML for different UAV-fluctuations.}
\vspace{-3mm}
\label{pdf}
\end{figure}

Fig. \ref{pdf} illustrates the PDF of the GML under two fluctuation scenarios; Scenario 1: $(\sigma_z,\sigma_{\theta_i},\sigma_{\phi_i})\!=\!(0.5\tilde{a}, 10^{-5}\ \text{rad},\\ 10^{-5}\ \text{rad})$, and Scenario 2: $(\sigma_z,\sigma_{\theta_i},\sigma_{\phi_i})\!=\!(0.6\tilde{a}, 10^{-4}\ \text{rad}, 10^{-4}\ \text{rad})$. The analytical model in \eqref{pdfh} shows good agreement with the empirical histogram based on \eqref{br} with only a slight deviation at higher fluctuations, consistent with the underlying assumptions. As expected, increased fluctuations degrade the channel quality. Asymptotically, higher fluctuations yield smaller $\frac{1}{m\beta_{\mathrm{g}}}$: it decreases from 2.54 to 0.72 for the QP profile and from 0.05 to 0.01 for the LP profile between the two scenarios. Particularly, when $\frac{1}{m\beta_{\mathrm{g}}}\!<\!1$, $\lim_{h_{\mathrm{irs}} \to 0} f_{h_{\mathrm{irs}}}(h_{\mathrm{irs}})\!=\! \infty$, which is consistent with the heavy tail observed in Fig. \ref{pdf} for the QP profile in Scenario 2. However, in Scenario 1, the tail is significantly shorter as $\frac{1}{m\beta_{\mathrm{g}}^{QP}}=2.54\!\!>\!\!1$, resulting in a faster PDF decay as $h_{\mathrm{irs}}$ decreases. The LP profile is expected to exhibit a qualitatively similar behavior, which becomes more pronounced as $h_{\mathrm{irs}} \to 0$.
We denote by $\alpha_{\mathrm{g}}^{QP}$ and $\alpha_{\mathrm{g}}^{LP}$ the maximum received power on the lens in the absence of misalignment for the QP and LP profiles, respectively. While $\alpha_{\mathrm{g}}^{QP}$ is sensitive to the focal distance and can be optimized to exceed $\alpha_{\mathrm{g}}^{LP}$, such optimization is left for future work. 

\begin{figure}[h]
\vspace{-5pt}
\centering
\includegraphics[width=0.75
\linewidth]{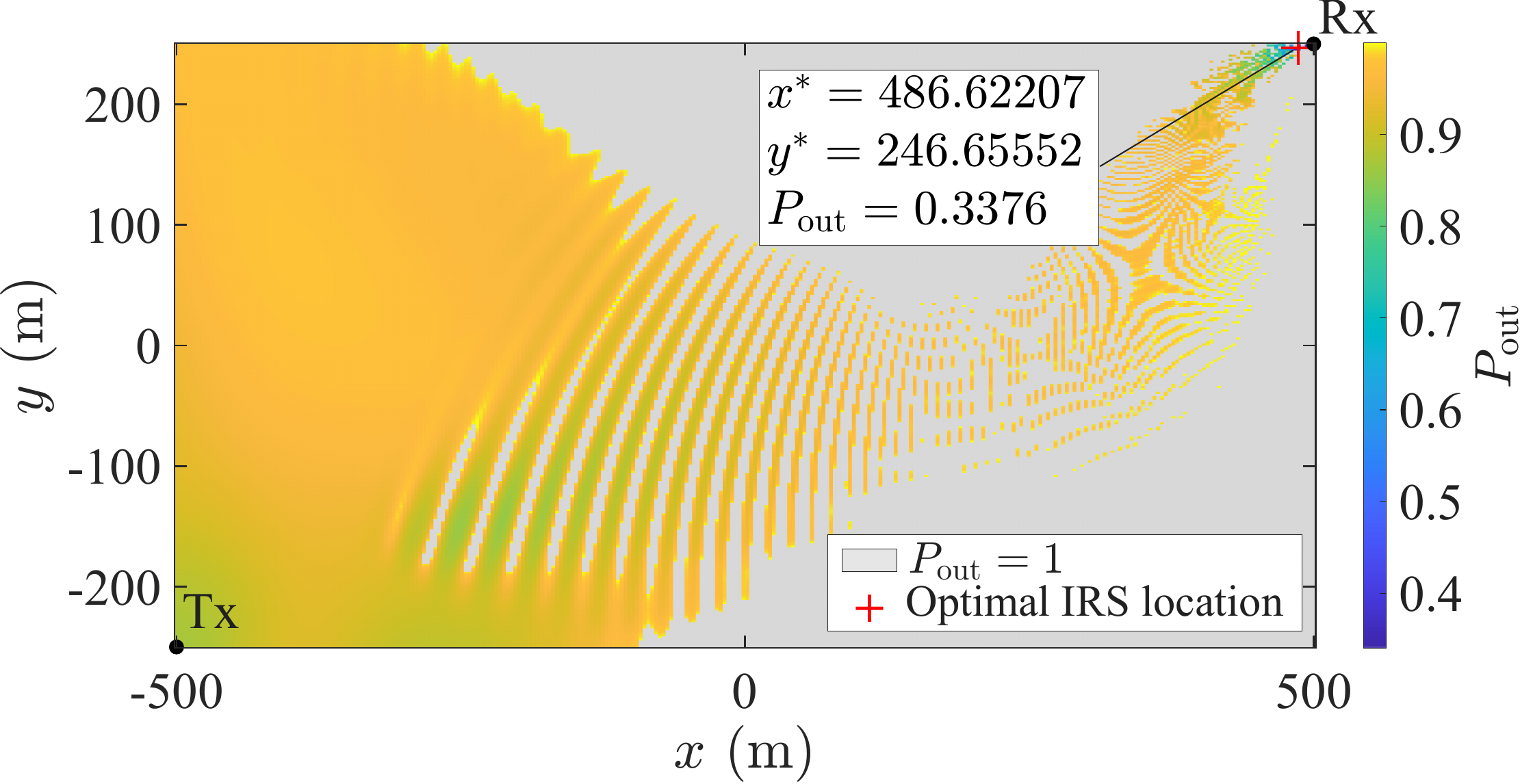}
\caption{Optimal location of the IRS for $\bar \rho= 20 \ \text{dB}$, $\rho_{\mathrm{th}}=0 \ \text{dB}$, and $(\sigma_z, \sigma_{\theta_i}, \sigma_{\phi_i})=(0.75a, 10^{-3},10^{-3})$.}
\vspace{-12pt}
\label{heat}
\end{figure}

In Fig. \ref{heat}, we provide insights into the optimal IRS position $(x^*,y^*)$ w.r.t. the $x_ty_tz_t$-coordinate system determined by minimizing the outage probability caused by beam misalignment for the QP profile. The Tx and Rx are located at $(-500,-250) \ \text{m}$ and $(500,250) \ \text{m}$, respectively, w.r.t. the $x_ty_tz_t$-coordinate system and the UAV-mounted IRS is at $H\!\!=\!\!100\ \text{m}$ above the $x_ty_t$-plane. Given the transmit signal-to-noise ratio (SNR) and the SNR threshold, denoted by $\bar{\rho}$ and $\rho_{\mathrm{th}}$, respectively, a numerical search is carried out over the region $\{(x,y)\!:\!-500\!\leq \!x \!\leq 500, \,-250 \!\leq \!y \!\leq\!250\}$, to minimize the outage probability given by 
\begin{align}
P_{\mathrm{out}} 
&= \Pr\!\left(\rho \leq \rho_{\mathrm{th}}\right)
= \Pr\!\left(\zeta^2 h_p^2 h_{\mathrm{irs}}^2 \bar{\rho} \le \rho_{\mathrm{th}}\right) \nonumber \\
&= F_{h_{\mathrm{irs}}}\!\left(\frac{1}{\zeta h_p}\sqrt{\frac{\rho_{\mathrm{th}}}{\bar{\rho}}}\right),
\quad 0 < \frac{1}{\zeta h_p}\sqrt{\frac{\rho_{\mathrm{th}}}{\bar{\rho}}} \le \alpha_{\mathrm{g}},
\label{pout}
\end{align}
where $\rho$ denotes the SNR at the receiver, and $F_{h_{\mathrm{irs}}}\!\left(\cdot\right)$ is the cumulative distribution function (CDF) of $h_{\mathrm{irs}}$ given by 
{\begin{equation}
F_{h_{\mathrm{irs}}}(h)
=1 \!-\! \frac{1}{\Gamma(n)}
\gamma\!\left(n,\frac{1}{m\beta_{\mathrm{g}} }\ln\!\left(\frac{\alpha_{\mathrm{g}}}{h}\right)\right), \quad 0 < h \leq \alpha_{\mathrm{g}}. 
\end{equation}}
Here, $\gamma(\cdot)$ denotes the lower incomplete Gamma function.
Furthermore, as indicated in \eqref{pout}, the condition $0 \!< \!\sqrt{\frac{\rho_{\mathrm{th}}}{\bar{\rho}}} \!\leq \!\zeta h_p \alpha_{\mathrm{g}}$ must be satisfied for $P_{\mathrm{out}}\!\!<\!\!1$. This imposes a constraint on any candidate $(x,y)$, as both $h_p$ and $\alpha_{\mathrm{g}}$ depend on the IRS position. 
The optimal IRS position (red $\!+\!$ in Fig. \ref{heat}) achieves an outage probability $P_{\mathrm{out}}\!=\! 0.34$ for $(\sigma_z, \sigma_{\theta_i}, \sigma_{\phi_i})\!=\! (0.75a, 10^{-3}\ \text{rad}, 10^{-3}\ \text{rad})$ and lies near the Rx at $(486.6, 246.7)\ \text{m}$ for the considered simulation scenario. As shown in Fig. \ref{heat}, the outage probability is uniformly high $(P_{\mathrm{out}}\approx 0.9)$ throughout most of the region surrounding the Tx, although slightly lower values can be observed in its direct proximity. In contrast, significantly lower values are achieved when the IRS is deployed in the direct vicinity of the Rx. This behavior highlights a key trade-off: while the IRS-induced beam focusing mitigates divergence, it also increases sensitivity to misalignment. Therefore, placing the IRS closer to the receiver, within the feasible region, reduces the propagation distance after reflection, thereby reducing the probability of large deviations from the center of the lens. Moreover, in the immediate vicinity of the optimal IRS position near the Rx, the outage probability exhibits sharp variations, going rapidly from $P_{\mathrm{out}}\!\approx\!0.9$ to values below $0.6$ over distances of only a few meters. This behavior indicates that even small deviations from the optimal IRS location can result in performance degradation. These observations highlight the importance of accurate UAV-mounted IRS positioning for practical FSO-assisted IRS deployments, particularly under UAV-induced fluctuations.
The local minima in the outage probability arise due to the operation in the Fresnel regime \cite{ajam2022modeling}, where diffraction induces oscillations in the received power on the lens ($\alpha_{\mathrm{g}}$), which are directly reflected in the outage performance. 
\vspace{-2mm}

\vspace{-1mm}
\section{Conclusion}\label{conc}
This paper developed a statistical channel model for the GML of an FSO link assisted by a UAV-mounted IRS. Using the Huygens-Fresnel principle, we derived general expressions for both the incident and received electric fields for any location of the transceiver, a closed-form expression for the conditional GML, and a tractable closed-form PDF for the GML under Gaussian angular and vertical UAV fluctuations, valid for both quadratic and linear phase-shift profiles. MC simulations confirmed the analysis and showed that, for the studied scenario, the optimal IRS location lies close to the Rx, since shortening the post-reflection path mitigates misalignment sensitivity more than the associated increase in geometric loss degrades performance.
\vspace{-3mm}
\section*{Appendix~A: Proof of Lemma~1} \label{app:lemma1}
The point $\mathbf{r}\!=\!(x,y,0)$ in the IRS coordinate system is first expressed in the LS  coordinate system as follows
{$\mathbf{r}_\ell \!=\!(x_\ell, y_\ell, z_\ell)\!=\!\mathbf{R}_{y_\ell}^{\top}\!\left(\frac{\pi}{2} - \theta_i\right)\, \mathbf{R}_{z}^{\top}(\phi_i)\, \mathbf{r} + \mathbf{c}_1$}, where {\small$\mathbf{c}_1\!=\!(0,0,d_1)$}, and {\small$
\small\mathbf{R}_y(\phi)\!=\!
{\begin{pmatrix}
\cos(\phi) & 0 & -\sin(\phi) \\
0 & 1 & 0 \\
\sin(\phi) & 0 & \cos(\phi)
\end{pmatrix}}
\quad \text{and} \\ \quad
\mathbf{R}_z(\phi) =
{\begin{pmatrix}
\cos(\phi) & \sin(\phi) & 0 \\
-\sin(\phi) & \cos(\phi) & 0 \\
0 & 0 & 1 \end{pmatrix}}$} are the counter-clockwise rotation matrices by angle $\phi$ around the $y$- and $z$-axes, respectively. We obtain $x_\ell=\sin(\theta_i)\cos(\phi_i)x-\sin(\theta_i)\sin(\phi_i)y$, $y_\ell=\sin(\phi_i)x+\cos(\phi_i)y$,  and $z_\ell=d_1-\cos(\theta_i)\cos(\phi_i)x+\cos(\theta_i)\sin(\phi_i)y$. Assuming $d_1\!\gg x_\ell,\; \!y_\ell$, we approximate $z_\ell \! \approx \! d_1$ in $w(\cdot)$, $R(\cdot)$, and $\tan^{-1}(\cdot)$ in \eqref{eq:Eli} \cite{ajam2022modeling}. Then, substituting $x_\ell$ and $y_\ell$ in \eqref{eq:Eli} by their respective expressions in terms of $x$, $y$, $\theta_i$, and $\phi_i$, we obtain (\ref{E_in}). $\xi_{\mathrm{in}}=\sqrt{|\sin(\theta_i)|}$ is derived according to the law of energy conservation \cite{ajam2022modeling},  \cite{thorne2017modern}.
\vspace{-2mm}
\section*{Appendix~B: Proof of Theorem~1}
\label{app:theorem1}
By using the Taylor approximation of $\lVert \mathbf{r}_o - \mathbf{r} \rVert$, given by  \eqref{ro}, in the exponential part of \eqref{E_r}, and \cite[Eq.~(2.33-1)]{gradshteyn2014table} to solve the integral w.r.t. $x$, we obtain the following expression for $E_\mathrm{r}(\mathbf{r}_o)$ 
\begin{align}\small
E_\mathrm{r}(\mathbf{r}_o)
&= C_\mathrm{r}
\int_{-\frac{L_y}{2}}^{\frac{L_y}{2}}
\int_{-\frac{L_x}{2}}^{\frac{L_x}{2}}
e^{-b_x x^2-b_y y^2-jka_xx-jka_yy-b_{xy}xy}
\,\mathrm{d}x\,\mathrm{d}y
\nonumber\\
&=
\frac{C_\mathrm{r}}{2}\sqrt{\frac{\pi}{b_x}}
\int_{-\frac{L_y}{2}}^{\frac{L_y}{2}}
\Bigg[
\operatorname{erf}\!\left(
\frac{\sqrt{b_x}L_x}{2}
+\frac{jka_x+b_{xy}y}{2\sqrt{b_x}}
\right)
-\operatorname{erf}\!\left(
-\frac{\sqrt{b_x}L_x}{2}
+\frac{jka_x+b_{xy}y}{2\sqrt{b_x}}
\right)
\Bigg]
\nonumber\\[-1mm]
&\qquad\times
e^{-b_y y^2-jka_y y
+\frac{(jka_x+b_{xy}y)^2}{4b_x}}
\,\mathrm{d}y,
\label{Erpr}
\end{align}
where $C_\mathrm{r}\!=\!\frac{C_i\xi_{\mathrm{in}}\xi_q}{j\lambda d_2}
e^{j\psi_0}$, {$a_x=\phi_x
-\cos(\theta_i)\cos(\phi_i)
-\frac{x_o}{d_2}$}, 
{$a_y = \phi_y
+\cos(\theta_i)\sin(\phi_i)
\!-\!\frac{y_o}{d_2}$}, $b_x=\frac{1}{w_x^2(d_1)}+ j k\left(
\frac{1-c_1^2d_2^2}{2d_2}
+\frac{1}{2R_x(d_1)} \\
+\phi_{x^2}\right)$, $b_y=\frac{1}{w_y^2(d_1)}
+j k\left(\frac{1-c_2^2d_2^2}{2d_2}
+\frac{1}{2R_y(d_1)}
+\phi_{y^2}\right)$, $b_{xy} = w_{xy}^{-2}(d_1) + jk\big(R_{xy}^{-1}(d_1) - c_1c_2d_2 + \phi_{xy}\big)$, $c_1=\frac{\cos(\phi_r)\cos(\theta_r)}{d_2}$, $c_2=\frac{\sin(\phi_r)\cos(\theta_r)}{d_2}$, $c_3=\frac{\cos(\phi_r)\sin(\theta_r)}{d_2}$, $c_4=\frac{\sin(\phi_r)}{d_2}$, $c_5=\frac{\sin(\phi_r)\sin(\theta_r)}{d_2}$, and $c_6=\frac{\cos(\phi_r)}{d_2}$. Subsequently, we replace {$y\!=\!\frac{L_y}{4}$} in $\operatorname{erf}(\cdot)$, assuming $|\frac{b_{xy} L_y}{2 \sqrt{b_x}}|\!\ll\!1$ holds. Then, we use the relation between the representation of a point on the lens in the IRS coordinate system $\mathbf{r}_o=(x_o, y_o,z_o)$ and in the lens coordinate system $\mathbf{r}_r=(x_r,x_r,0)$, given by
{$\mathbf{r}_o 
= \mathbf{R}_{z}(-\phi_r)\, \mathbf{R}_{y_r}\!\left(\theta_r - \frac{\pi}{2}\right) (\mathbf{r}_r+\mathbf{c}_2)$}, where {$\mathbf{c}_2=(0,0,d_2)$}, to obtain $x_o=\cos(\phi_r)\sin(\theta_r)x_r-\sin(\phi_r)y_r+d_2\cos(\phi_r)\cos(\theta_r)$, $y_o=\sin(\phi_r)\sin(\theta_r)x_r+\cos(\phi_r)y_r+d_2\sin(\phi_r)\cos(\theta_r)$, and $z_o=-\cos(\theta_r)x_r+d_2\sin(\theta_r)$, respectively. We substitute $x_o$ and $y_o$ in \eqref{Erpr} and use the following approximations $\lVert\mathbf{r}_o \rVert \simeq d_2$, $z_o \simeq d_2\sin(\theta_r)$,
$\frac{x_o^2}{d_2^3}\simeq 
\frac{\cos^2(\phi_r)\cos^2(\theta_r)}{d_2}$, $\frac{y_o^2}{d_2^3} \simeq 
\frac{\sin^2(\phi_r)\cos^2(\theta_r)}{d_2}$, $\frac{x_o y_o}{d_2^3}\simeq 
\frac{\cos^2(\theta_r)\cos(\phi_r)\sin(\phi_r)}{d_2}$. Finally, we solve the integral w.r.t. $y$ using \cite[Eq.~(2.33-1)]{gradshteyn2014table}.
\vspace{-2mm}
\section*{Appendix~C: Proof of Theorem~2} \label{app:theorem2}
By substituting $x_r\!-\!u\!\approx\!\frac{\tilde{a}}{2}$ in $\operatorname{erf}(\cdot)$ in \eqref{hirs1}, similarly to \cite{sipani2024irs}, we obtain the simplified expression 
\begin{align}
h_{\mathrm{irs}}
&\!=\!
\frac{\pi|\tilde C|^2|C_1|^2|C_2|^2}
{8\eta P_t\sqrt{\!\rho_y\tilde{\rho}_x}}\!
\Big[\!\operatorname{erf}(\sqrt{\rho_y}\tilde{a} \!+\!\tfrac{\rho_{xy}\tilde{a}}{4\sqrt{\rho_y}})\!
-\!\operatorname{erf}(-\sqrt{\rho_y}\tilde{a} \!
+\!\tfrac{\rho_{xy}\tilde{a}}{4\sqrt{\rho_y}})\!
\Big]\Big[\!\operatorname{erf}(\sqrt{\tilde{\rho}_x}(\tilde{a}+u))
\!-\!\operatorname{erf}(-\sqrt{\tilde{\rho}_x}(\tilde{a}-u))\Big]\\ \nonumber
&\stackrel{(a)}{=} \frac{\pi|\tilde C|^2|C_1|^2|C_2|^2}
{4\eta P_t\sqrt{\!\rho_y\tilde{\rho}_x}}\sqrt{\frac{\tilde{\rho}_x}{\pi}}\Big[\!\operatorname{erf}(\sqrt{\rho_y}\tilde{a} \!+\!\tfrac{\rho_{xy}\tilde{a}}{4\sqrt{\rho_y}})\!
-\!\operatorname{erf}(-\sqrt{\rho_y}\tilde{a} \!
+\!\tfrac{\rho_{xy}\tilde{a}}{4\sqrt{\rho_y}})\!
\Big]\int_{-(\tilde{a}-u)}^{(\tilde{a}+u)} e^{-\tilde{\rho}_x x^{2}}\!\, dx.
\end{align}

Using $
2\sqrt{\frac{c}{\pi}}\!\int_{p}^{q} e^{-c x^{2}}\!\, dx
=\!\operatorname{erf}(q\sqrt{c})\!- \!\operatorname{erf}(p\sqrt{c})$ in $(a)$
and the Taylor expansion of the exponential function, $h_{\mathrm{irs}}$ can be written as
{\small$h_{\mathrm{irs}}
\!=\!\sum_{k=0}^{\infty}\!A_{2k} u^{2k}$}. Then, by equating the first two terms of $h_{\mathrm{irs}}$ to $\alpha_{\mathrm{g}} e^{-\beta_{\mathrm{g}} u^2}$, where $\beta_{\mathrm{g}} \!>\! 0$, we obtain \eqref{hirs2}.
\section*{Appendix~D: Proof of Theorem~3}\label{app:theorem3}
The expressions of $b_{r,x}$ and $b_{r,y}$ in terms of $\epsilon_z$, $\epsilon_{\theta_r}$, and $\epsilon_{\phi_r}$, derived in Section~\ref{misal}, using first-order Taylor expansions around the mean values $\mu_z$, $\mu_{\theta_r}$, and $\mu_{\phi_r}$, are given by $b_{r,x}=p_x-\mu_z \cot(\mu_{\theta_r})\cos(\mu_{\phi_r})+\mu_z \cot(\mu_{\theta_r})\sin(\mu_{\phi_r})\epsilon_{\phi_r}+\frac{\mu_z\cos(\mu_{\phi_r})}{\sin(\mu_{\theta_r})^2}\epsilon_{\theta_r}-\cot(\mu_{\theta_r})\cos(\mu_{\phi_r})\epsilon_z$ and $b_{r,y}=p_y+\mu_z \cot(\mu_{\theta_r})\sin(\mu_{\phi_r})+\mu_z \cot(\mu_{\theta_r})\\ \cos(\mu_{\phi_r})\epsilon_{\phi_r}-\frac{\mu_z\sin(\mu_{\phi_r})}{\sin(\mu_{\theta_r})^2}\epsilon_{\theta_r}+\cot(\mu_{\theta_r})\sin(\mu_{\phi_r})\epsilon_z$, respectively.$b_{r,x}$ and $b_{r,y}$ as sums of Gaussian RVs, so they are also Gaussian distributed. However, $b_{r,x}$ and $b_{r,y}$ are correlated since they are both functions of $\epsilon_{\theta_r}$, $\epsilon_{\phi_r}$, and $\epsilon_z$. Hence, the joint distribution of $b_{r,x}$ and $b_{r,y}$ is a bivariate Gaussian distribution.
$\tilde{\mathbf{b}}_{r}=(b_{r,x},b_{r,y})
\sim \mathcal{N}\!\left(
\boldsymbol{\mu},
\mathbf{\Sigma}
\right),$
where $\boldsymbol{\mu}\!=\!(\mu_1,\mu_2)$ with $\mu_1\!=p_x-\mu_z \cot(\mu_{\theta_r})\cos(\mu_{\phi_r})$ and $\mu_2=p_y+\mu_z \cot(\mu_{\theta_r})\sin(\mu_{\phi_r})$. Let $\mathbf{\Sigma}= \mathbf{U} \mathbf{\Lambda} \mathbf{U}^{\top}$
be the eigenvalue decomposition of $\mathbf{\Sigma}$, where $\mathbf{\Lambda}$ is a diagonal matrix
with elements $\lambda_1$ and $\lambda_2$, and $\mathbf{U}$ is a unitary matrix. 
We can show that
$\tilde{\mathbf{b}}_{r}= \boldsymbol{\omega}\mathbf{U}^{\top},$ 
where 
$\boldsymbol{\omega}=(\omega_x,\omega_y) \sim 
\mathcal{N}
\left(
\mathbf{U}^{\top}\boldsymbol{\mu},
\mathbf{\Lambda}
\right)$. Hence, $u= \sqrt{\tilde{\mathbf{b}}_{r}\tilde{\mathbf{b}}_{r}^{\top}}\!=\!\sqrt{\boldsymbol{\omega}\mathbf{U}^{\top}\!\mathbf{U}\boldsymbol{\omega}^{\top}}\!=\!\sqrt{\omega_x^2 \!+\! \omega_y^2}$,
where $\omega_x$ and $\omega_y$ are independent Gaussian RVs
with different means and different variances. Therefore, $u$ follows a Beckmann distribution given by
\begin{equation}
f_u(u)
\!=\!\frac{u}{2\pi \sigma_{\omega_x}\sigma_{\omega_y}}\!
\int_{0}^{2\pi}\!
e^
{-\frac{(u\cos\phi - \mu_1)^2}{2\sigma_{\omega_x}^2}
- \frac{(u\sin\phi-\mu_2)^2}{2\sigma_{\omega_y}^2}} \mathrm{d}\phi,
\quad u\!>\!0, \nonumber
\end{equation} where $\sigma_{\omega_x}^2\!=\!\lambda_1$ and $\sigma_{\omega_y}^2\!=\!\lambda_2$ are the eigenvalues of the covariance matrix $\mathbf{\Lambda}$. As demonstrated in \cite{10816714}, the PDF of $r=u^2$ can be accurately approximated by a Gamma distribution given by
\begin{align}
f_r(r)
&=
\frac{1}{\Gamma(k)\theta^k}
r^{k-1}e^{-\frac{r}{\theta}},
\qquad r>0, \nonumber
\end{align}
where $\theta
=\frac{\operatorname{Var}(r)}{\mathbb{E}\{r\}} 
\ \text{and} \ 
k=\frac{\mathbb{E}^2\{r\}}{\operatorname{Var}(r)} \ \text{with}
\operatorname{Var}(r)
=\mathbb{E}\{r^2\}
-\mathbb{E}^2\{r\}$.

Finally, using \eqref{relpdfs}, we obtain $f_{h_\mathrm{irs}}(\cdot)$ in \eqref{pdfh}.

\bibliographystyle{IEEEtran}
\bibliography{sample}

@article{ajam2022modeling,
  author    = {H. Ajam and others},
  title     = {Modeling and Design of {IRS}-Assisted Multilink {FSO} Systems},
  journal   = {IEEE Trans. Commun.},
  volume    = {70},
  number    = {5},
  pages     = {3333--3349},
  year      = {2022}
}

@book{thorne2017modern,
  author    = {K. S. Thorne and R. D. Blandford},
  title     = {Modern Classical Physics: Optics, Fluids, Plasmas, Elasticity, Relativity, and Statistical Physics},
  publisher = {Princeton Univ. Press},
  address   = {Princeton, NJ, USA},
  year      = {2017}
}

@inproceedings{10001121,
  author    = {H. Ajam and others},
  title     = {Power Scaling Law for Optical {IRS}s and Comparison With Optical Relays},
  booktitle = {Proc. IEEE Global Commun. Conf. (GLOBECOM)},
  pages     = {1527--1533},
  address   = {Rio de Janeiro, Brazil},
  year      = {2022},
  doi       = {10.1109/GLOBECOM48099.2022.10001121}
}

@book{gradshteyn2014table,
  author    = {I. S. Gradshteyn and I. M. Ryzhik},
  title     = {Table of Integrals, Series, and Products},
  edition   = {8th},
  publisher = {Academic Press},
  address   = {Amsterdam, The Netherlands},
  year      = {2014}
}

@article{ajam2023optical,
  author    = {H. Ajam and others},
  title     = {Optical {IRS}s: Power Scaling Law, Optimal Deployment, and Comparison With Relays},
  journal   = {IEEE Trans. Commun.},
  volume    = {72},
  number    = {2},
  pages     = {954--970},
  year      = {2023}
}

@article{10816714,
  author    = {E. Zedini and others},
  title     = {A Novel Approach to Approximating Generalized Pointing Errors Modeled by {Beckmann} Distribution in {FSO} Communication Systems},
  journal   = {IEEE Open J. Commun. Soc.},
  volume    = {6},
  pages     = {727--741},
  year      = {2025},
  doi       = {10.1109/OJCOMS.2024.3523304}
}

@article{10821003,
  author    = {F. Tarhouni and others},
  title     = {Free Space Optical Mesh Networks: A Survey},
  journal   = {IEEE Open J. Commun. Soc.},
  volume    = {6},
  pages     = {642--655},
  year      = {2025},
  doi       = {10.1109/OJCOMS.2025.3525468}
}

@article{8869705,
  author    = {W. Saad and others},
  title     = {A Vision of 6{G} Wireless Systems: Applications, Trends, Technologies, and Open Research Problems},
  journal   = {IEEE Netw.},
  volume    = {34},
  number    = {3},
  pages     = {134--142},
  year      = {2020},
  doi       = {10.1109/MNET.001.1900287}
}

@article{safari2008relay,
  author    = {M. Safari and M. Uysal},
  title     = {Relay-Assisted Free-Space Optical Communication},
  journal   = {IEEE Trans. Wireless Commun.},
  volume    = {7},
  number    = {12},
  pages     = {5441--5449},
  year      = {2008}
}

@article{najafi2021intelligent,
  author    = {M. Najafi and others},
  title     = {Intelligent Reflecting Surfaces for Free-Space Optical Communication Systems},
  journal   = {IEEE Trans. Commun.},
  volume    = {69},
  number    = {9},
  pages     = {6134--6151},
  year      = {2021}
}

@article{ndjiongue2022design,
  author    = {A. R. Ndjiongue and others},
  title     = {Design of a Power Amplifying-{RIS} for Free-Space Optical Communication Systems},
  journal   = {IEEE Wireless Commun.},
  volume    = {28},
  number    = {6},
  pages     = {152--159},
  year      = {2022}
}

@article{ndjiongue2021analysis,
  author    = {A. R. Ndjiongue and others},
  title     = {Analysis of {RIS}-Based Terrestrial-{FSO} Link Over {GG} Turbulence With Distance and Jitter Ratios},
  journal   = {J. Lightw. Technol.},
  volume    = {39},
  number    = {21},
  pages     = {6746--6758},
  year      = {2021}
}

@article{11122651,
  author    = {H. T. Le and others},
  title     = {Leveraging {RIS} and {QoS}-Aware Transmission Rate for Energy-Efficient {FSO} Non-Terrestrial Networks},
  journal   = {IEEE Trans. Aerosp. Electron. Syst.},
  volume    = {61},
  number    = {6},
  pages     = {16528--16540},
  year      = {2025},
  doi       = {10.1109/TAES.2025.3597911}
}

@article{malik2022performance,
  author    = {S. Malik and others},
  title     = {Performance Analysis of a {UAV}-Based {IRS}-Assisted Hybrid {RF/FSO} Link With Pointing and Phase Shift Errors},
  journal   = {J. Opt. Commun. Netw.},
  volume    = {14},
  number    = {4},
  pages     = {303--315},
  year      = {2022}
}

@article{trinh2025optical,
  author    = {P. V. Trinh and others},
  title     = {Optical {RIS}s Improve the Secret Key Rate of Free-Space {QKD} in {HAP}-to-{UAV} Scenarios},
  journal   = {IEEE J. Sel. Areas Commun.},
  year      = {2025}
}

@inproceedings{sipani2024irs,
  author    = {J. Sipani and others},
  title     = {{IRS}-Assisted {UAV}-Based {FSO} System: Modeling Approach for Hovering {UAV}},
  booktitle = {Proc. IEEE 100th Veh. Technol. Conf. (VTC-Fall)},
  pages     = {1--5},
  year      = {2024}
}

@article{11106762,
  author    = {F. Tarhouni and others},
  title     = {Performance Analysis of {SAGIN} From the Relay Perspective: A Spherical Stochastic Geometry Approach},
  journal   = {IEEE Trans. Aerosp. Electron. Syst.},
  volume    = {61},
  number    = {6},
  pages     = {16313--16326},
  year      = {2025},
  doi       = {10.1109/TAES.2025.3594693}
}

@article{najafi2020statistical,
  author    = {M. Najafi and others},
  title     = {Statistical Modeling of the {FSO} Fronthaul Channel for {UAV}-Based Communications},
  journal   = {IEEE Trans. Commun.},
  volume    = {68},
  number    = {6},
  pages     = {3720--3736},
  year      = {2020}
}

@article{sipani2023modeling,
  author    = {J. Sipani and others},
  title     = {Modeling and Design of {IRS}-Assisted {FSO} System Under Random Misalignment},
  journal   = {IEEE Photon. J.},
  volume    = {15},
  number    = {4},
  pages     = {1--13},
  year      = {2023}
}

\end{document}